\documentclass[superscriptaddress,secnumarabic,
amssymb,amsmath,nobibnotes,aps,prd,showkeys,showpacs,nofootinbib]{revtex4}%
\usepackage{graphicx}
\usepackage{epsf}
\usepackage{bm}
\usepackage{amsmath}
\usepackage{amsfonts}
\usepackage{amssymb}
\usepackage{epstopdf}
\usepackage{natbib}
\usepackage{color}
\usepackage{float}
\providecommand{\U}[1]{\protect\rule{.1in}{.1in}}

\newcommand{\be}{\begin{equation}}
\newcommand{\ee}{\end{equation}}

\newcommand{\mincir}{\raise
-3.truept\hbox{\rlap{\hbox{$\sim$}}\raise4.truept\hbox{$<$}\ }}
\newcommand{\magcir}{\raise
-3.truept\hbox{\rlap{\hbox{$\sim$}}\raise4.truept\hbox{$>$}\ }}

\begin{document}

\title{The problem of calculating  the $\beta$-Bogoliubov coefficient in Non-Oscillating models }

\author{Jaume  de Haro}
\email{jaime.haro@upc.edu}
\affiliation{Departament de Matem\`atiques, Universitat Polit\`ecnica de Catalunya, Diagonal 647, 08028 Barcelona, Spain}

\author{Llibert Arest\'e Sal\'o}
\email{l.arestesalo@qmul.ac.uk}
\affiliation{School of Mathematical Sciences, Queen Mary University of London, Mile End Road, London, E1 4NS, United Kingdom}



\begin{abstract}
The calculation of the Bogoliubov coefficients is a key piece to obtain the reheating temperature of the Universe. In all cases this calculation is performed either in toy models where some derivative of the potential is discontinuous at some points or by making some approximations to the model. The result of these calculations is applied to more realistic models without checking if they really apply to this realistic model because the exact calculation of the Bogoliubov coefficients for a viable model, which is usually depicted by a smooth potential, requires  a very complicated numerical calculation. Here we want to point out the difficulties that one encounters when trying to compute numerically these coefficients for a smooth potential without making approximations, which could lead to completely different results. 
\end{abstract}

\vspace{0.5cm}

 \par
\maketitle
\section{Introduction}

As was stated by Alan Guth in his seminal paper \cite{guth}, a viable inflationary model needs a reheating mechanism to match inflation with the hot Big Bang scenario. In standard inflation, where the potential contains a deep well, the reheating mechanism is quite simple: at the end of the inflationary period the inflaton field falls in the well and starts oscillating, thus releasing its energy and creating particles, which reheats the universe after thermalization .

\

However, after the discovery of the current cosmic acceleration \cite{riess,perlmutter}, some models without a deep well were proposed, the so-called Non-Oscillating models, such as Quintessential Inflation \cite{pv}, which tried to unify inflation with the current cosmic acceleration, 
where the inflaton field evolves monotonically. For these models, there are basically two mechanisms to produce particles which will reheat the universe. The first one is the so-called {\it gravitational particle production} \cite{parker, ford} and the other one is the  {\it instant preheating} \cite{fkl0,fkl}. In the former case, when  the reheating is due to the production  of very light particles, the calculation of the $\beta$-Bogoliubov coefficients -the main ingredient to calculate the energy density of the produced particles- is performed only for toy models with some discontinuities in some derivative of the scale factor (see for instance \cite{damour,giovannini}), obtaining that the energy density of these produced particles is proportional to the scale of inflation to the fourth power. This result is, of course, applied to more smooth models but without checking its validity, for example in \cite{pv}. When the reheating is due to the gravitational production of superheavy particles, which have to decay into lighter ones, one encounters the same situation: The $\beta$-Bogoliubov coefficients can only be calculated analytically for potentials with a discontinuity in some of its derivatives. The problem is that the more realistic models depend on a smooth potential and then the calculation of these coefficients requires, as we will see, very complicated numeric calculations. Finally, dealing with instant preheating, to calculate analytically the $\beta$-Bogoliubov coefficients one has to perform some approximation to obtain a parabolic problem which has a well-known analytical solution, but it is not justified that the correct value does resemble the analytical one, obtained with these approximations.

\

Therefore, in this work we try to explain systematically some different ways to find numerically, for smooth potentials,  the exact value -without unjustified approximations- of the Bogoliubov coefficients and the problems  we have encountered. Basically, we have checked that two different ways could lead to different results, meaning that our numerical calculations are not reliable at all.  Moreover, we are sure that this  problem is related with the background  which, for smooth potentials,  has to be as well calculated numerically. More precisely, the problem comes from the scale factor, which due to the inflationary period grows more than 25 orders, and then its numerical calculation needs also an interpolation which does not guarantee at all its real value.

\

The present work is organized as follows: In Section II we start presenting the well-known diagonalization method, obtaining the equations that satisfy the Bogoliubov coefficients. Once we have these equations we consider an approximation that does not affect the final result of our calculations.
Section III is devoted to the gravitational particle production. Starting with toy models, we calculate analytically the value of the $\beta$-Bogoliubov coefficient and we show that this analytical value does not coincide with any of the values obtained with our numerical methods. We also explain and perform the numerical calculations for a smooth potential, obtaining results that do not seem reliable as we explain in the text. 
In Section IV we deal with {\it instant preheating}, explaining the approximations made to obtain the analytical expression of the Bogoliubov coefficients, and pointing out that a numerical calculation without these approximations is needed to validate the analytic results. The importance of the Bogoliubov coefficients is stressed in Section V, where we show the expression of the reheating temperature and its bounds as a function of these coefficients. Finally we present our conclusions in the last section. We have included as well an appendix where we explain more thoroughly the algorithm that has been used for obtaining the numerical results and the methods that have been employed.

\section{The diagonalization method}

Given a massive quantum scalar field   conformally coupled to gravity, 
namely $\chi$, the Klein-Gordon (K-G) equation in the Fourier space,  which is satisfied by the $\chi$-modes in the flat Friedmann-Lema{\^\i}tre-Robertson-Walker (FLRW) spacetime, is given by 
\begin{eqnarray}\label{kg1}
\chi_{ k}''(\tau)+\omega^2_k(\tau) \chi_{ k}(\tau)=0,
\end{eqnarray}
where the prime denotes the derivative with respect to the conformal time $\tau$ and $\omega_k(\tau)$ is the time-dependent frequency, which
in the case of {\it gravitational particle production} \cite{parker,ford} is given by 
\begin{eqnarray}\omega_k^2(\tau)=k^2+a^2(\tau)m_{\chi}^2,\end{eqnarray}
being $m_{\chi}$ the heavy mass of the produced particles and $a$ the scale factor. When particles are produced via {\it Instant Preheating} \cite{fkl0,fkl}, the frequency is given  by
\begin{eqnarray}\omega_k^2(\tau)=k^2+a^2(\tau)[m_{X}^2+ g^2(\varphi-\varphi_{kin})^2],\end{eqnarray}
where $m_X$ stands for the bare mass of the produced particles, which is not related with $m_{\chi}$, $\varphi$ is the inflaton field and $\varphi_{kin}$ its value at the beginning of kination  (the regime where practically all energy is kinetic).
 
 \
 
 As usual, the modes that define the
 vacuum state at a given initial time $\tau_i$ are the ones that minimize the energy density, so they  must
 satisfy  the conditions
 \begin{eqnarray}\label{vacuum}
 \chi_{k}(\tau_i)=
 \frac{1}{\sqrt{2\omega_k(\tau_i)}}e^{-i\int^{\tau_i} \omega_k(\bar\eta)d\bar\eta}, \quad
 \chi_{ k}'(\tau_i)=
-i \omega_k(\tau_i)\chi_{ k}(\tau_i), \end{eqnarray}
and thus, 
 the vacuum expectation value of the energy density  will be  given by \cite{Bunch}

\begin{eqnarray}\label{vacuum-energy}
\langle\rho(\tau)\rangle\equiv \langle 0| \hat{\rho}(\tau)|0 \rangle=
\frac{1}{4\pi^2a^4(\tau)}\int_0^{\infty} k^2dk \left(   |\chi_{ k}'(\tau)|^2+ \omega^2_k(\tau) |\chi_{ k}(\tau)|^2-  \omega_k(\tau)        \right),
\end{eqnarray}
where, in order to obtain a finite energy density, one has to  subtract the energy density of the zero-point oscillations of the vacuum 
$\frac{1}{(2\pi)^3a^4(\tau)}\int d^3k  \frac{1}{2} \omega_k(\tau)$.

\

Following the method developed in \cite{Zeldovich} (see also Section $9.2$ of \cite{gmmbook}),  we will write
the modes as follows,
\begin{eqnarray}\label{zs}
\chi_{k}(\tau)= \alpha_k(\tau)\frac{e^{-i\int^{\tau} \omega_k(\bar\tau)d\bar\tau}}{\sqrt{2\omega_k(\tau)}}+
\beta_k(\tau)\frac{e^{i\int^{\tau} \omega_k(\bar\tau)d\bar\tau}}{\sqrt{2\omega_k(\tau)}},\end{eqnarray}
where $\alpha_k(\tau)$ and $\beta_k(\tau)$ are the time-dependent Bogoliubov coefficients.
Now, imposing that the modes satisfy   the condition
\begin{eqnarray}
\chi_{k}'(\tau)= -i\omega_k(\tau)\left(\alpha_k(\tau)\frac{e^{-i\int^{\tau} \omega_k(\bar\tau)d\bar\tau}}{\sqrt{2\omega_k(\tau)}}-
\beta_k(\tau)\frac{e^{i\int^{\tau} \omega_k(\bar\tau)d\bar\tau}}{\sqrt{2\omega_k(\tau)}}\right),\end{eqnarray}
one can show that   the Bogoliubov coefficients must satisfy the system 
\begin{eqnarray}\label{Bogoliubovequation}
\left\{ \begin{array}{ccc}
\alpha_k'(\tau) &=& \frac{\omega_k'(\tau)}{2\omega_k(\tau)}e^{2i\int^{\tau} \omega_k(\bar\tau)d\bar\tau}\beta_k(\tau)\\
\beta_k'(\tau) &=& \frac{\omega_k'(\tau)}{2\omega_k(\tau)}e^{-2i\int^{\tau}\omega_k(\bar\tau)d\bar\tau}\alpha_k(\tau),\end{array}\right.
\end{eqnarray}
in order for the  expression (\ref{zs}) to be a solution of the equation (\ref{kg1}).

\

Finally, inserting (\ref{zs}) into the expression for the vacuum energy density (\ref{vacuum-energy}), 
and taking into account that the Bogoliubov coefficients satisfy the equation $|\alpha_k(\tau)|^2- |\beta_k(\tau)|^2=1$,
one finds the following diagonalized form of the energy density \cite{Zeldovich},
\begin{eqnarray}\label{vacuum-energy1}
\langle\rho(\tau)\rangle= \frac{1}{2\pi^2a^4(\tau)}\int_0^{\infty} k^2\omega_k(\tau)|\beta_k(\tau)|^2 dk,
\end{eqnarray}
where it is important to notice that $|\beta_k(\tau)|^2$ encodes the vacuum polarization effects and also the production of real particles, which are only produced when the adiabatic evolution breaks. In non-oscillating models this happens during the phase transition from the end of inflation to the beginning of kination (a period where all the energy density is kinetic, i.e., the potential energy is negligible), and fortunately the polarization effects disappear immediately after the beginning of kination, when the value of $|\beta_k(\tau)|$ stabilizes to a value which we will denote by $|\beta_k|$, and thus, only encodes the production of real particles.

\

A final remark is in order: Once one has obtained the scale factor -in Section 3 we explain how to obtain it-, instead of solving numerically \eqref{Bogoliubovequation}, another way to calculate the value of $\beta_k$ (the value of the $\beta$-Bogoliubov coefficient when it stabilizes) is to solve numerically the equation (\ref{kg1}), with initial conditions (\ref{vacuum}), for example,  at the horizon crossing, i.e.,  
\begin{eqnarray}
 \chi_{k}(\tau_*)=
 \frac{1}{\sqrt{2\omega_k(\tau_*)}} , \quad
 \chi_{ k}'(\tau_*)=
-i \omega_k(\tau_*)\chi_{ k}(\tau_*), \end{eqnarray}
where the star denotes that the quantities are evaluated at the horizon crossing. 
This means that at that moment the quantum field is at the vacuum. In fact, as we will see later, for the relevant modes the quantum field is in the adiabatic vacuum at the horizon crossing, and it does not matter if one chooses  an early time  for the initial conditions, because the relevant modes continue in the adiabatic vacuum. 

\

Then, after the beginning of kination the Bogoliubov coefficients stabilize and the $k$-mode has the simple form
\begin{eqnarray}
\chi_{k}(\tau)= \alpha_k\phi_k(\tau)+
\beta_k \bar{\phi}_k(\tau),\end{eqnarray}
where we have used the notation
\begin{eqnarray}
\phi_k(\tau)= \frac{e^{-i\int^{\tau} \omega_k(\bar\tau)d\bar\tau}}{\sqrt{2\omega_k(\tau)}},
\end{eqnarray}
and $\bar{\phi}_k(\tau)$ is the conjugate of ${\phi}_k(\tau)$.

\

Then, using the Wronskian $W[f,g]\equiv fg'-f'g$, one has
\begin{eqnarray}
\beta_k=\frac{W[\chi_k,\phi_k]}{W[\bar{\phi}_k,\phi_k]}.
\end{eqnarray}

\

This is the first method that we present to calculate the $\beta$-Bogoliubov coefficient, and it is
 used in \cite{hashiba} to obtain its numerical results. However, as we will explain later there are other forms to obtain $\beta_k$.

\subsection{Approximation}
Using once again the relation $|\alpha_k(\tau)|^2-|\beta_k(\tau)|^2=1$, and since $\beta_k(\tau)$ has to be very small, one can take the well-justified approximation  $\alpha_k(\tau)\cong 1$, and (\ref{Bogoliubovequation}) becomes the simpler equation 
\begin{eqnarray}
\beta_k'(\tau) =\frac{\omega_k'(\tau)}{2\omega_k(\tau)}e^{-2i\int^{\tau}\omega_k(\bar\tau)d\bar\tau},
\end{eqnarray}
whose solution is given by 
\begin{eqnarray}\label{A}
\beta_k(\tau) =\int_{\tau^*}^{\tau}\frac{\omega_k'(\tilde{\tau})}{2\omega_k(\tilde{\tau})}
e^{-2i\int^{\tilde{\tau}}_{\tau_*}\omega_k(\bar\tau)
d\bar\tau}d\tilde{\tau},
\end{eqnarray}
where we have assumed that $\beta_k(\tau)$ vanishes at the horizon crossing, which also means that at the horizon crossing  the quantum field is at the vacuum. This is a second way to calculate the Bogoliubov coefficients.

\

The problem is that this oscillatory integral seems difficult to be computed.
For this reason, in order to calculate this coefficient we make another  approximation \cite{ema} to simplify its calculation:
\begin{enumerate}
    \item When $k\leq a_*m_{\chi}$, one can make the approximation $\omega_k(\tau)\cong a(\tau)m_{\chi}$ obtaining
    \begin{eqnarray}
    \beta_k(\tau)\cong \int_{t_*}^{t}\frac{H(t)}{2}e^{-2im_{\chi}(t-t_*)}dt, \end{eqnarray}
    where $H$ is the Hubble rate and $t$ denotes the cosmic time.

\item When $k\geq a_*m_{\chi}$, we define  $\tau_k$ such that
$k=a(\tau_k)m_{\chi}$, and we make the approximation $\omega_k(\tau)\cong k$ for
$\tau<\tau_k$ and $\omega_k(\tau)\cong a(\tau)m_{\chi}$ when $\tau>\tau_k$, obtaining 
    
        \begin{eqnarray}
        \beta_k(\tau)\cong \int_{\tau_*}^{\tau_k}\frac{a(\tau)a'(\tau)m_{\chi}^2}{2k^2}e^{-2ik(\tau-\tau_*)}d\tau+e^{-2ik(\tau_k-\tau_*)}\int_{t_k}^{t}\frac{H(t)}{2}e^{-2im_{\chi}(t-t_k)}dt.
    \end{eqnarray}

\end{enumerate}

However, the problem of this approximation is that the error made doing it could be greater than the value of the $\beta$-Bogoliubov coefficient. So, in practice this approximation does not seem viable.

\

Finally, a third method comes from the 
 important observation  that the modulus of the solution (\ref{A}) is also the modulus of the solution of the simple linear differential equation 
\begin{eqnarray}\label{B}
y_k'-2i\omega_k(\tau)y_k=\frac{\omega'_k(\tau)}{2\omega_k(\tau)},
\end{eqnarray}
with initial condition $y_k(\tau_*)=0$,  because one has $|\beta_{k}(\tau)|=|y_k(\tau)|$ by taking into account the formula of variation of parameters for first order differential equations.

\subsection{The energy density of produced particles}
Let $\tau_s$ be the moment when $|\beta_k(\tau)|$ stabilizes, and as we have already stated we will denote this value by
 $|\beta_k|$. Then,  after the beginning of kination, the energy density of the real particles   will be
\begin{eqnarray}\label{vacuum-energy1}
\langle\rho(\tau)\rangle= \frac{1}{2\pi^2a^4(\tau)}\int_0^{\infty} k^2\omega_k(\tau)|\beta_k|^2 dk.
\end{eqnarray}

Focusing our analysis on the case of gravitational particle production, i.e., when $\omega_k(\tau)=\sqrt{k^2+a^2(\tau)m^2_{\chi}}$, 
and considering superheavy particles satisfying $m_{\chi}\gg H_*$ (here $H_*$ denotes the scale of inflation, i.e., the value of the Hubble rate at the horizon crossing), one expects that the modes that really contribute to the energy density satisfy 
  $k\leq a_{kin}m_{\chi}$ where the sub-index $kin$ denotes that the quantities are evaluated at the beginning of kination. 
  We expect that these modes are the only relevant ones because this is what happens
  for the toy models that we will study in the next section, and the physical reason is that the real particles are produced during the phase transition from the end  of inflation to the beginning of kination, so the modes satisfying $k\gg a_{kin}m_{\chi}$ do not feel the gravity because they satisfy the K-G equation $\chi_k''+k^2\chi_k=0$ which corresponds to the Minkowski modes and where particles are not produced. 
  
  \
  
  For these modes  one can make the approximation 
$\omega_k(\tau)\cong a(\tau)m_{\chi}$, and thus, the energy density of produced particles becomes
\begin{eqnarray}\label{vacuum-energy2}
\langle\rho(\tau)\rangle\cong \frac{m_{\chi}}{2\pi^2a^3(\tau)}\int_0^{a_{kin}m_{\chi}} k^2|\beta_k|^2 dk.
\end{eqnarray}

\

Here it is important to remark that the modes satisfying $k\leq a_*m_{\chi}$ do not contribute to the energy density of produced particles. Effectively,  the contribution of those modes is
\begin{eqnarray}
\frac{m_{\chi}}{2\pi^2a^3(\tau)}\int_0^{a_*m_{\chi}} k^2|\beta_k|^2 dk
\leq \frac{m_{\chi}}{2\pi^2a^3(\tau)}\int_0^{a_*m_{\chi}} k^2 dk
=\frac{m_{\chi}^4}{6\pi^2}\left( \frac{a_*}{a(\tau)}\right)^3,
\end{eqnarray}
where we have used that $|\beta_k|\leq 1$.

Now, taking into account that $a(\tau)\geq a_*e^{N}$, where $N$ is the number of efolds from the horizon exit to the end of inflation, being satisfied that $N>60$ for models with a kination phase, we conclude that the contribution of these modes is less than 
\begin{eqnarray}
\frac{m_{\chi}^4}{6\pi^2}e^{-3N},
\end{eqnarray}
which is in practice a negligible energy density.

\

Using the same reasoning one can see that the modes satisfying $k\geq a_{kin}H_*$ are in the adiabatic vacuum at the horizon crossing. Effectively, we have to evaluate $\frac{\omega_k'(\tau_*)}{\omega_k^2(\tau_*)}$ whose value is given by
\begin{eqnarray}
\frac{\omega_k'(\tau_*)}{\omega_k^2(\tau_*)}=\frac{a_*^3H_*m_{\chi}^2}{\omega_k^3(\tau_*)}\leq 
\frac{m_{\chi}^2}{H_*^2}\left( \frac{a_*}{a_{kin}}\right)^3\leq \frac{m_{\chi}^2}{H_*^2}e^{-3N}\ll 1,
\end{eqnarray}
where we have used that $\omega_k(\tau_*)\cong k\leq a_{kin}H_*$.

\

A final remark is in order: When one deals with superheavy particles the energy density (\ref{vacuum-energy1}) has to evolve as matter, meaning that the evolution of the energy density is
\begin{eqnarray}
\langle \rho(\tau)\rangle =\langle \rho_{kin}\rangle\left(\frac{a_{kin}}{a(\tau)} \right)^3,
\end{eqnarray}
where $\langle \rho_{kin}\rangle$ is the value of (\ref{vacuum-energy1}) at the beginning of kination. This important result, as we will see in the next section, could be checked studying simple toy models.

\section{Gravitational particle production}
\subsection{Toy models}
To warm up, here  we study some simple models containing discontinuities. 
First of all, we consider the following scale factor (see for instance \cite{hashiba}),
\begin{eqnarray}
a(t)=\left\{\begin{array}{ccc}
a_*e^{H_{inf}(t-t_*)}&\mbox{when}& t\leq t_{kin}\\
& &\\
a_{kin}\left( \frac{t}{t_{kin}}\right)^{1/3}&\mbox{when}& t\geq t_{kin},\end{array}
\right.
\end{eqnarray}
where $a_*=1$,  $H_{inf}=10^{-6} M_{pl}$,  $t_{kin}=\frac{1}{3H_{inf}}$, $t_*=t_{kin}-\frac{65}{H_{inf}}$ and $a_{kin}=a(t_{kin})=e^{65}$.

For this scale factor,  which depicts a sudden phase transition from de Sitter to kination, one can calculate analytically the $\beta_k$ Bogoliubov coefficient
using the formula (\ref{A}).
Since the derivative of the Hubble rate is discontinuous at the time $\tau_{kin}$, integrating two times  by parts and disregarding the terms in $\tau_*$ and $\tau_s$ (we consider only relevant modes, that is, satisfying $k\gg a_*m_{\chi}$, then at the horizon crossing the corresponding terms are negligible because the adiabatic condition $\omega_k'/\omega_k^2\ll 1$ is strongly satisfied) one can deduce that the leading term is given by
\begin{eqnarray}\label{b}
|\beta_k|\cong \frac{3a_{kin}^4H^2_{inf}m_{\chi}^2}{8\omega_k^4(\tau_{kin})},
\end{eqnarray}
which for $k\ll a_{kin}m_{\chi}$ is of the order 
\begin{eqnarray}
|\beta_k|\sim \frac{3H_{inf}^2}{8m_{\chi}^2}.
\end{eqnarray}

Then, for example, choosing  a heavy  mass of the order of $m_{\chi}\sim 10^{15}$ GeV one has
\begin{eqnarray}
|\beta_k|^2\sim 5\times 10^{-12}.
\end{eqnarray}

Therefore, for modes in the range $a_*m_{\chi}\ll k\ll a_{kin}m_{\chi}$ one would have to obtain numerically the same result. On the contrary, if the numerical results differ from the analytical ones, this means that the numerical method used does not work.

\

Note also that one can calculate analytically the energy density inserting the Bogoliubov coefficient (\ref{b}) in (\ref{vacuum-energy2}),  and using the formula
\begin{eqnarray}
\int_0^{\infty}\frac{dx}{(1+x^2)^{n+1}}=\frac{2n-1}{2n}\int_0^{\infty}\frac{dx}{(1+x^2)^{n}}
\end{eqnarray}
one gets 
\begin{eqnarray}
\langle \rho(\tau)\rangle= \frac{9H_{inf}^4}{16^3\pi}\left(\frac{a_{kin}}{a(\tau)} \right)^3\sim 10^{-3}H_{inf}^4\left(\frac{a_{kin}}{a(\tau)} \right)^3,
\end{eqnarray}
which coincides with the result obtained in \cite{hashiba} and  shows that this energy density evolves like matter.

\

This result shows that the relevant modes, the ones that contribute to the energy density, are the ones that satisfy $k\leq a_{kin}m_{\chi}$ because the contribution of these modes is equal to 
\begin{eqnarray}
\frac{9H_{inf}^4}{128\pi^2}\left(\frac{\pi}{64}-\frac{1}{48}\right)\left(\frac{a_{kin}}{a(\tau)} \right)^3,
\end{eqnarray} which is of the same order than the expression obtained above. This is consistent, as we have already expalined,  with the fact that for $k\gg a_{kin}m_{\chi}$ at the beginning of kination $\omega_k(\tau_{kin})\cong k$, and thus, the K-G equation is given at that moment by
$\chi_k''+k^2\chi_k=0$, i.e., they are in the Minkowski vacuum and  consequently they do not feel the gravity. Thus, they do not produce particles.

\

The second toy model is given by the potential 
\begin{eqnarray}\label{quadratic}
V(\varphi)=\left\{\begin{array}{ccc}
\frac{1}{2}m^2\varphi^2 &\mbox{for}& \varphi\leq  0\\
0 &\mbox{for}& \varphi\geq 0,
\end{array}\right.
\end{eqnarray}
where $m\cong 5\times 10^{-6} M_{pl}$, $\varphi_*\cong -\frac{2\sqrt{2}}{\sqrt{1-n_s}}M_{pl}$ with $n_s\cong 0.96$, and
$H_*=\sqrt{\frac{V(\varphi_*)}{3M_{pl}^2}}\cong \sqrt{\frac{m^2\varphi_*^2}{6M_{pl}^2}}.$

\

Now, the second derivative of the potential is discontinuous, which means that the third derivative of the Hubble rate is discontinuous.

\

In that case, for  $a_*m_{\chi}\ll k \ll a_{kin}m_{\chi}$, integrating (\ref{A}) four times by parts, disregarding the boundary terms because for the relevant modes the adiabatic condition $\omega_k'/\omega_k^2\ll 1$ is strongly satisfied at the boundary points,  and using the Raychaudhuri equation $\dot{H}=-\frac{1}{ 2M_{pl}^2}\dot{\varphi}^2$, after a cumbersome calculation one gets
\begin{eqnarray}
|\beta_k|\cong \frac{3a_{kin}^6m^4m_{\chi}^2}{32\omega_k^6(\tau_{kin})}\cong\frac{3m^4}
{32m_{\chi}^4},
\end{eqnarray}

which for $m_{\chi}\sim 10^{15}$ GeV leads to 
\begin{eqnarray}
|\beta_k|^2\sim 4\times 10^{-18}.
\end{eqnarray}

\

Effectively, integrating four times by parts one gets

\begin{eqnarray}
|\beta_{k}|\cong \frac{|\omega_k''''(\tau_{kin}^{-})-\omega_k''''(\tau_{kin}^{+})|}{32\omega^5_{kin}},
\end{eqnarray}
where  the beginning of kination $\tau_{kin}$ corresponds to the value of the scalar field $\varphi=0$.

And, since  $|\omega_k''''(\tau_{kin^-})- \omega_k''''(\tau_{kin^+})|= \frac{a^6_{kin}|\dddot{H}(\tau_{kin}^{-})-\dddot{H}(\tau_{kin}^{+})|m_{\chi}^2}{\omega_{kin}}$,
one gets
\begin{eqnarray}
|\beta_{k}|\cong \frac{a_{kin}^6|\dddot{H}(\tau_{kin}^{-})-\dddot{H}(\tau_{kin}^{+})|m_{\chi}^2}{32\omega^6_{kin}}.
\end{eqnarray}

Now, taking the temporal derivative of the Raychaudhuri equation we have $\ddot{H}=\dot{\varphi}\ddot{\varphi}/M_{pl}^2$, and thus
\begin{eqnarray}
|\dddot{H}(\tau_{kin}^{-})-\dddot{H}(\tau_{kin}^{+})|= \frac{\dot{\varphi}_{kin}}{M_{pl}^2}
|\dddot{\varphi}(\tau_{kin}^{-})-\dddot{\varphi}(\tau_{kin}^{+})|.
\end{eqnarray}

On the other hand, from the conservation equation we have
\begin{eqnarray}
|\dddot{\varphi}(\tau_{kin}^{-})-\dddot{\varphi}(\tau_{kin}^{+})|\cong V_{\varphi\varphi}(0^+)\dot{\varphi}_{kin}=m^2\dot{\varphi}_{kin},
\end{eqnarray}
so
\begin{eqnarray}
|\dddot{H}(\tau_{kin}^{-})-\dddot{H}(\tau_{kin}^{+})|= \dot{\varphi}_{kin}^2m^2/M_{pl}^2.
\end{eqnarray}

Finally, at the end of inflation we have
$\varphi_{END}=\sqrt{2}M_{pl}$ and $H_{END}^2= \frac{3V(\varphi_{END})/2}{3M_{pl}^2}= m^2/2$. Assuming that there is no substantial drop of energy between the end of inflation and the beginning of kination $m/\sqrt{2}=H_{END}\cong H_{kin}$ since during kination all energy is kinetic, one gets
$\dot \varphi_{kin}\cong \sqrt{3}mM_{pl}$, which leads to
\begin{eqnarray}
|\dddot{H}(\tau_{kin}^{-})-\dddot{H}(\tau_{kin}^{-})|\cong 3m^4.
\end{eqnarray}

\

Unfortunately for the model (\ref{quadratic}), as we can see in Figure 1, the numerical results differ in $8$ orders of magnitude with respect to the analytical ones.

\begin{figure}[ht]
\centering
\includegraphics[width=0.8\textwidth]{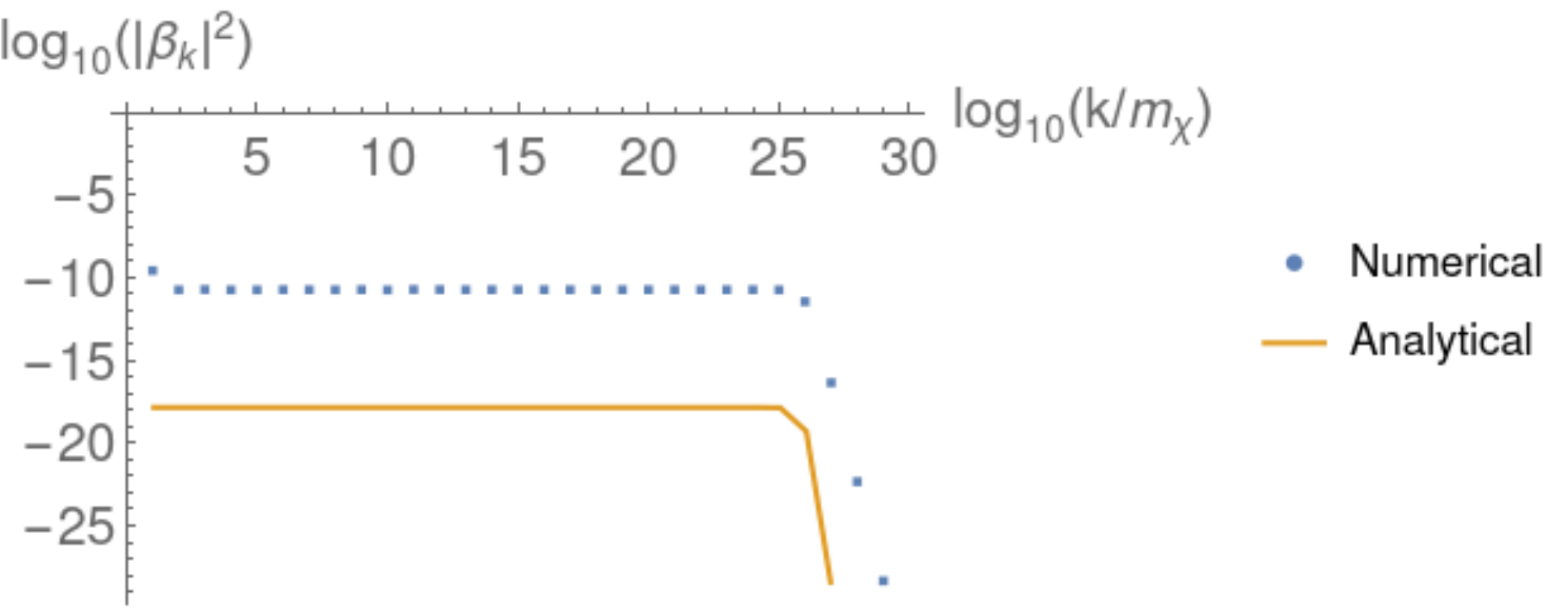}
\caption{Plot of the analytical and numerical values of $|\beta_k|^2$ for different values of $k$ obtained by solving equation \eqref{B} for $m_{\chi}=10^{15}$ GeV. We can see that for values of $k$ greater than $a_{kin}m_{\chi}$ (for this model we have checked numerically that choosing $a_*=1$ one has $a_{kin}\sim 10^{28}$) the square modulus of the $\beta$-Bogoliubov coefficient decreases very fast as happens with the analytical solution. However, there is a difference of around $8$ orders between both solutions.}
\label{fig:beta_simple}
\end{figure}

\

To conclude this section, one can also calculate the energy density after the beginning of kination for this model, obtaining 

\begin{eqnarray}
\langle \rho(\tau)\rangle= \frac{315 m^8}{8\pi\times 16^4 m_{\chi}^4}\left(\frac{a_{kin}}{a(\tau)} \right)^3\cong 9\times 10^{-8} m^4\left(\frac{a_{kin}}{a(\tau)} \right)^3,
\end{eqnarray}
for masses of the order $10^{15}$ GeV.

\


\subsection{Smooth models}
Given a non-oscillating potential, for example those which appear in Quintessential Inflation, in order to solve (\ref{Bogoliubovequation}) first of all one has to obtain the background, i.e., integrate  numerically the conservation equation for the inflaton field in terms of the cosmic time, namely
\begin{eqnarray}\label{conservation}
\ddot{\varphi}+3H\dot{\varphi}+V_{\varphi}=0,
\end{eqnarray}
where $H=\frac{1}{\sqrt{3}M_{pl}}\sqrt{\frac{\dot{\varphi}^2}{2}+V(\varphi)  }$, with initial conditions at the horizon crossing,
 i.e., when the pivot scales leaves the Hubble radius. Recall that in that moment
the system is in the slow-roll phase and, since this regime is an attractor, one only has to take initial conditions in the basin of attraction of the slow-roll solution, for example,
$\varphi=\varphi_*$ and $\dot{\varphi}=-\frac{V_{\varphi}(\varphi_*)}{3H_*}$, where the ``star" denotes that the quantities are evaluated
at the horizon crossing.

\

Once one has obtained the evolution of the background, and in particular the evolution of the Hubble rate, one computes the evolution of the scale factor, which is 
given by 
\begin{eqnarray}
a(t)=a_*e^{\int_{t_*}^t H(s)ds},
\end{eqnarray}
where the value of $a_*$ is arbitrary and can be chosen to be $a_*=1$.

\

\begin{figure}[ht]
\centering
\includegraphics[width=0.4\textwidth]{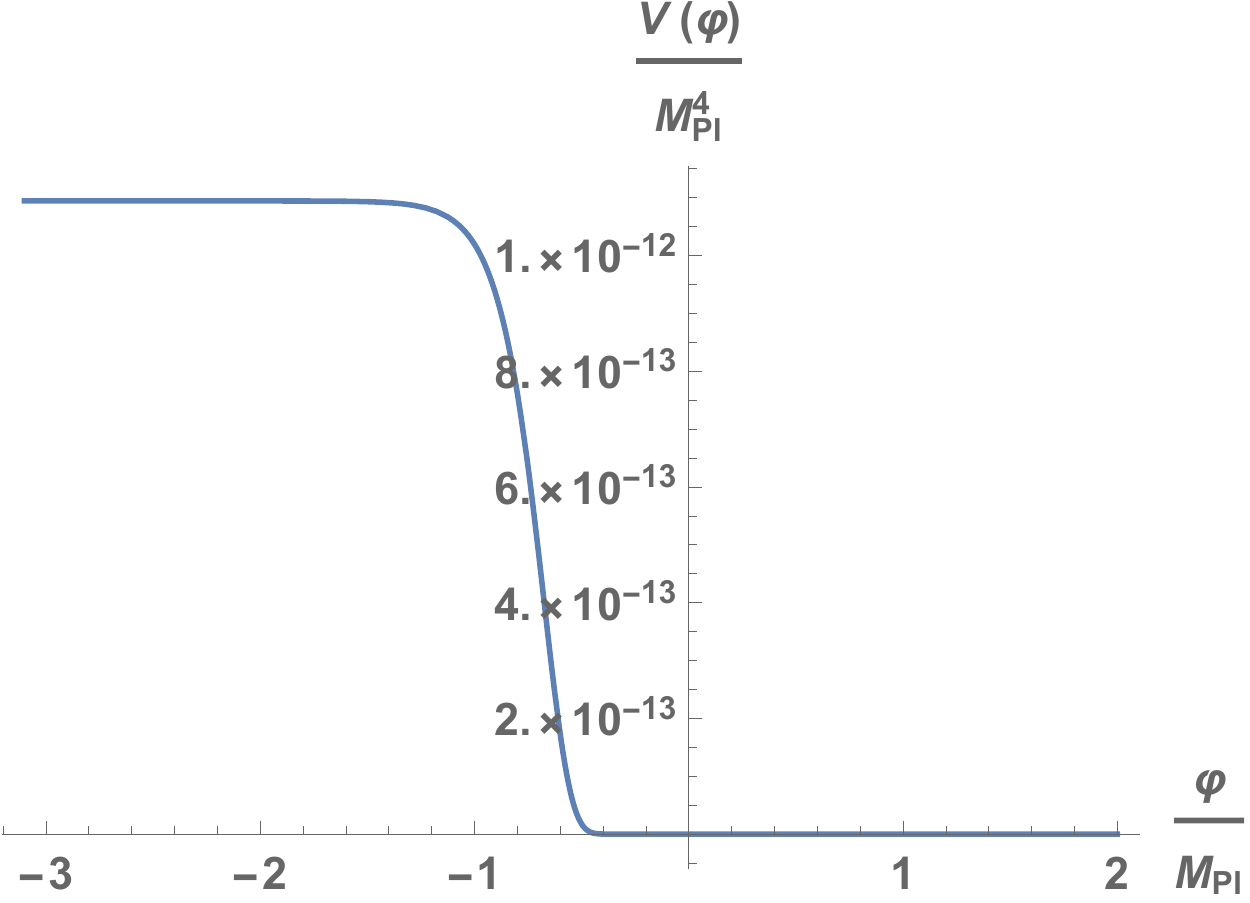}
\caption{{Plot of the Exponential $\alpha$-attractor potential, for $\alpha\sim 10^{-2}$, $n\sim 10^2$ and $\lambda\sim 10^{-66}$.}}
\label{fig:attr}
\end{figure}

More precisely, for the $\alpha$-attractor potential displayed in Figure \ref{fig:attr},
\begin{eqnarray}\label{alpha}
V(\varphi)=\lambda M_{pl}^4e^{-n\tanh\left(\frac{\varphi}{\sqrt{6\alpha}M_{pl}} \right)},
\end{eqnarray}
where $\lambda$, $\alpha$ and $n$ are dimensionless parameters which have to satisfy the following relation in order to match with the current observation data,
\begin{eqnarray}\label{parameters}
\frac{\lambda}{\alpha}e^{n}\sim 10^{-10} \qquad \mbox{and}\qquad \lambda e^{-n}\sim 10^{-120},
\end{eqnarray}
 one has to choose the values of $\alpha$, for example $\alpha=10^{-2}, 10^{-1}$
and $1$, and for a given value of $\alpha$ from equation (\ref{parameters}) one finds the corresponding values of $\lambda$ and $n$.

The value of $\varphi_*$ is obtained from the relations  (see for details	\cite{haro})
\begin{eqnarray}\label{slowroll}
\epsilon_*\hspace{-0.1cm}=\hspace{-0.1cm}\frac{n^2}{12\alpha}\frac{1}{\cosh^4\left( \frac{\varphi_*/M_{pl}}{\sqrt{6\alpha}}\right)}, \qquad  \eta_*\hspace{-0.1cm}
\cong\hspace{-0.1cm} -\frac{n}{3\alpha}\frac{1}{\cosh^2\left(\frac{\varphi_*/M_{pl}}{\sqrt{6\alpha}} \right)},
\end{eqnarray}
with $\varphi_*<0$. And using the relation 
\begin{eqnarray}\label{power}
n_s\cong 
1-6\epsilon_*+2\eta_*\cong
1-\frac{2}{N}, \qquad  r\cong 16\epsilon_*\cong\frac{12\alpha}{N^2},
\end{eqnarray}
one gets from the equation (\ref{power})
\begin{eqnarray}
\epsilon_*=\frac{3\alpha}{16}(1-n_s)^2,
\end{eqnarray}
which, together with the expression of $\epsilon_*$ given in (\ref{slowroll}), leads to
\begin{eqnarray}\label{a}
\cosh\left(\frac{\varphi_*}{\sqrt{6\alpha} M_{pl}}  \right)=
\sqrt{\frac{2n}{3\alpha(1-n_s)}},
\end{eqnarray}
whose solution is given by
\begin{eqnarray*}
\varphi_*=\sqrt{6\alpha}M_{pl}\ln\left(\sqrt{\frac{2n}{3\alpha(1-n_s)}}-\sqrt{\frac{2n}{3\alpha(1-n_s)}-1}
\right), \qquad n_s\cong 0.96.
\end{eqnarray*}

\

In the same way, from the formula of the power spectrum of scalar perturbations
\begin{eqnarray}
H_*^2=16\pi^2\epsilon_*10^{-9}M_{pl}^2,
\end{eqnarray}
and using  $\epsilon_*=
\frac{3\alpha}{16}(1-n_s)^2$, one gets the value of $H_*$,
\begin{eqnarray}
H_*\cong 6\sqrt{\alpha}\times 10^{-6}M_{pl}.
\end{eqnarray}

Now, taking into account that 
\begin{eqnarray}\label{alpha}
V_{\varphi}(\varphi_*)=-\frac{n\lambda M_{pl}^3}{\sqrt{6\alpha}}
\frac{e^{-n\tanh\left(\frac{\varphi_*}{\sqrt{6\alpha}M_{pl}} \right)}}{\cosh\left(\frac{\varphi_*}{\sqrt{6\alpha}M_{pl}} \right)}\cong-\frac{1}{2}
\sqrt{\frac{3\alpha}{2}}(1-n_s)e^{-n}M_{pl}^3,\end{eqnarray} where we have used the formula
(\ref{a}), we have
\begin{eqnarray}
\dot{\varphi}_*=-\frac{V_{\varphi}(\varphi_*)}{3H_*}\cong \frac{1}{12}
\sqrt{\frac{3}{2}}(1-n_s)10^6e^{-n}M_{pl}^2.
\end{eqnarray}


\

In Figure $3$ we have plotted the evolution of the scalar field $\varphi$. For $\alpha=10^{-2}$ we have obtained 
$\varphi_*\cong-1.7 M_{Pl}$, $\varphi_{END}\cong-0.89M_{Pl}$, $\varphi_{kin}\cong -0.54M_{Pl}$, where, as we have already used,  the sub-index ``END" denotes the end of inflation.

\begin{figure}[ht]
\includegraphics[width=0.48\textwidth]{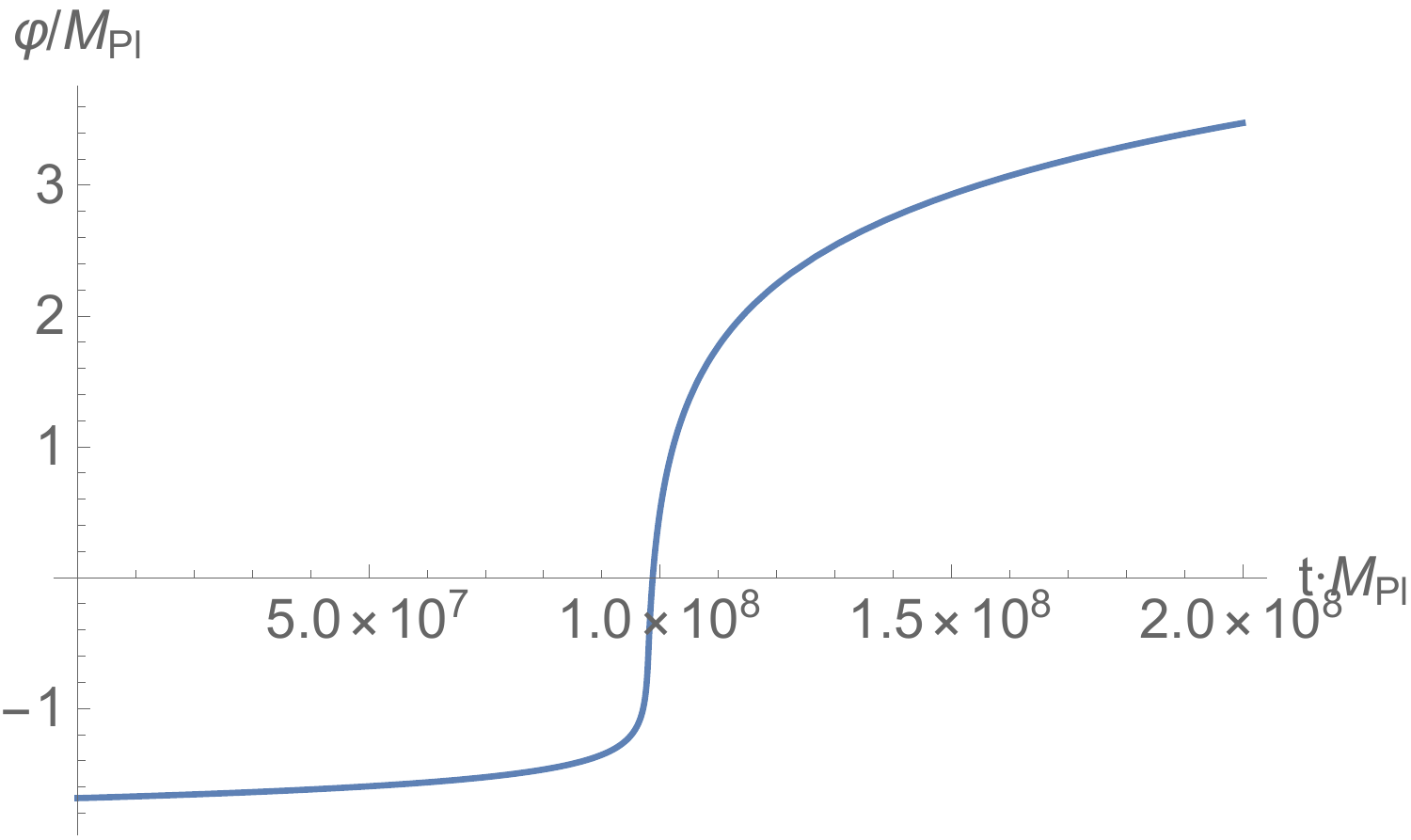}
\caption{Evolution  of the scalar field $\varphi$, when $\alpha\sim 10^{-2}$.}
\label{fig:phi_evolution}
\end{figure}

\begin{figure}[ht]
\includegraphics[width=0.48\textwidth]{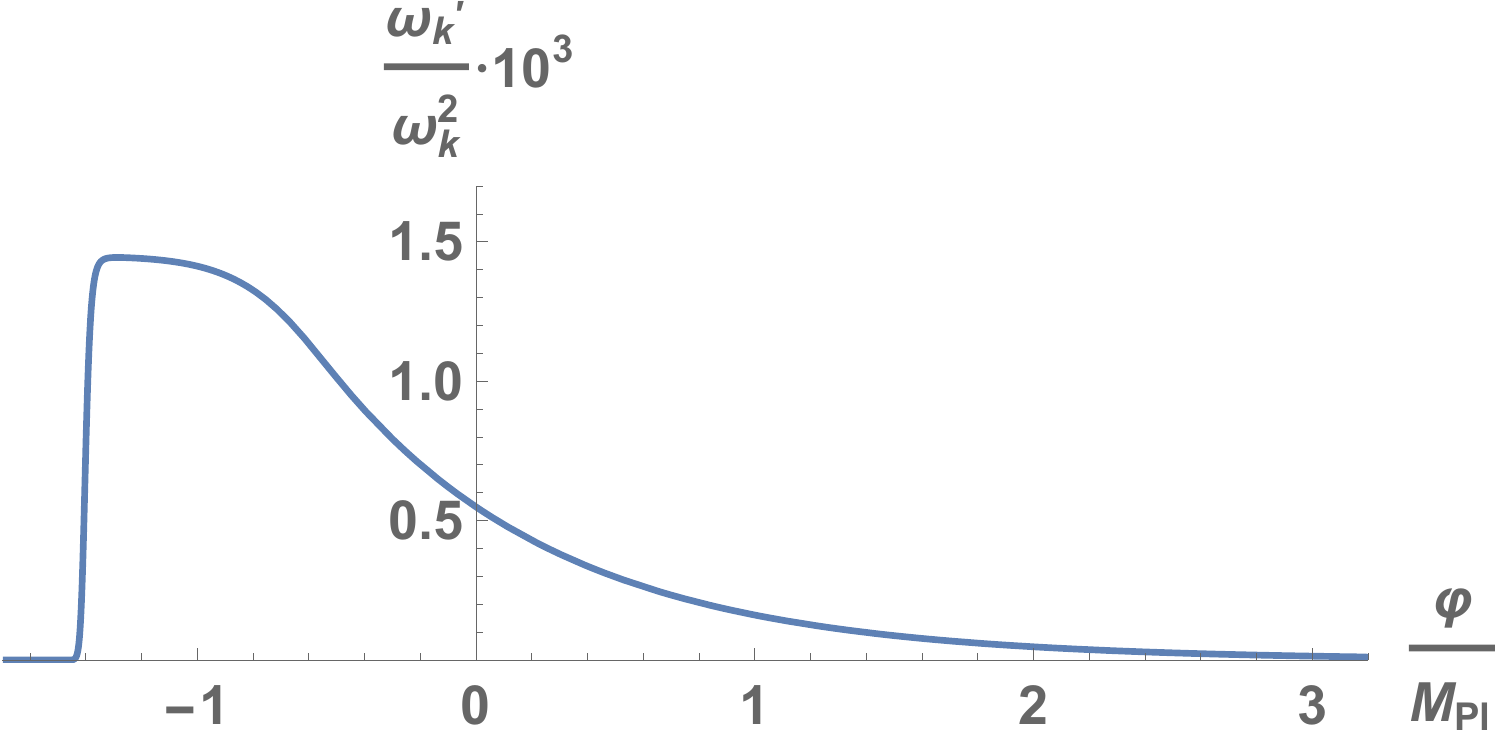}
\caption{Adiabatic evolution  for a heavy field with mass $m_{\chi}\sim 10^{15}$ GeV, when $\alpha\sim 10^{-2}$. Here we have used the value $k=a_{kin}H_{kin}$.
}
\label{fig:bogo}
\end{figure}

Then, in \cite{haro1,haro2}  the equation \eqref{Bogoliubovequation} is transformed into a second order differential equation, namely
\begin{eqnarray}
\left\{ \begin{array}{cc}
     &  \alpha_k''(\tau)=\alpha_k'(\tau)\left(\frac{\omega_k''(\tau)}{\omega_k'(\tau)} - \frac{\omega_k'(\tau)}{\omega_k(\tau)}+2i\omega_k(\tau) \right)+ \left(\frac{\omega_k'(\tau)}{2\omega_k(\tau)} \right)^2\alpha_k(\tau)\\
     & \beta_k''(\tau)=\beta_k'(\tau)\left(\frac{\omega_k''(\tau)}{\omega_k'(\tau)} - \frac{\omega_k'(\tau)}{\omega_k(\tau)}-2i\omega_k(\tau) \right)+ \left(\frac{\omega_k'(\tau)}{2\omega_k(\tau)} \right)^2\beta_k(\tau)
\end{array}\right..
\end{eqnarray}

Assuming that at the horizon crossing the quantum field is the vacuum, which is equivalent to taking the 
 initial conditions 
 $\alpha_k(\tau_*)=1$ and $\beta_k(\tau_*)=0$ leading  to $\alpha_k'(\tau_*)=0$, one can solve  the equation for $\alpha_k(\tau)$, which can be split into the real and imaginary parts in the following way,
\begin{eqnarray}\label{X}
\left\{ \begin{array}{cc}
     &  \alpha_{k,Re}''(\tau)=\alpha_{k,Re}'(\tau)\left(\frac{\omega_k''(\tau)}{\omega_k'(\tau)} - \frac{\omega_k'(\tau)}{\omega_k(\tau)}\right)-2\omega_k(\tau)\alpha_{k,Im}'(\tau) + \left(\frac{\omega_k'(\tau)}{2\omega_k(\tau)} \right)^2\alpha_{k,Re}(\tau)\\
     & \alpha_{k,Im}''(\tau)=\alpha_{k,Im}'(\tau)\left(\frac{\omega_k''(\tau)}{\omega_k'(\tau)} - \frac{\omega_k'(\tau)}{\omega_k(\tau)}\right)+2\omega_k(\tau)\alpha_{k,Re}'(\tau) + \left(\frac{\omega_k'(\tau)}{2\omega_k(\tau)} \right)^2\alpha_{k,Im}(\tau)
\end{array}\right.,
\end{eqnarray}
and then $|\beta_k(\tau)|^2=|\alpha_k(\tau)|^2-1$ because of the well-known conservation property of the Wronskian. This is the last method that we will present here. 

\

Numerical calculations have been done in the case of the exponential $\alpha$-attractor \cite{haro2}
\begin{eqnarray}\label{alpha}
V(\varphi)=\lambda M_{pl}^4e^{-n\tanh\left(\frac{\varphi}{\sqrt{6\alpha}M_{pl}} \right)},
\end{eqnarray}
where
by solving numerically (\ref{X}) we have
obtained that, for masses  $m_{\chi}\cong 10^{15}-10^{17}$ GeV and for a large range of values of $\alpha$,  the modes which are in the range 
$a_{kin}H_{kin}\lesssim k\lesssim a_{kin}m_{\chi}$ lead to
 values of $|\beta_k|^2$ of order $10^{-9}$ for $m_{\chi}\sim 10^{15}$ GeV and values of $|\beta_k|^2$ of order $10^{-10}$ for $m_{\chi}\sim 10^{16}-10^{17}$ GeV.

\

However, the numerical results obtained solving the exact solution (\ref{X}) (see Appendix \ref{app} for more details) do not seem very confident for many reasons:
\begin{enumerate}
    \item The method breaks for modes out of the range $a_{kin}H_{kin}\lesssim k\lesssim a_{kin}m_{\chi}$.
    \item It has been impossible to obtain numerically $\beta_k$ using the approximate formula (\ref{A}). If the results obtained solving (\ref{X}) are correct, it seems clear that one has to obtain the same ones using the integral form (\ref{A}).
    \item The same happens using the differential equation (\ref{B}). We have tried to obtain the same results solving (\ref{B}), but we have not been able to solve numerically this equation.
    
     \item The results solving numerically (\ref{X}) for the smooth potential (\ref{alpha}) lead to values of $|\beta_k|^2$ greater than the analytical ones obtained from the potential (\ref{quadratic}), whose second derivative has a discontinuity. This seems a contradiction.

    \item Using this method we cannot reproduce the analytical results obtained from the toy models, as one can see in Figure $5$.
    
    \begin{figure}[ht]
\centering
\includegraphics[width=0.8\textwidth]{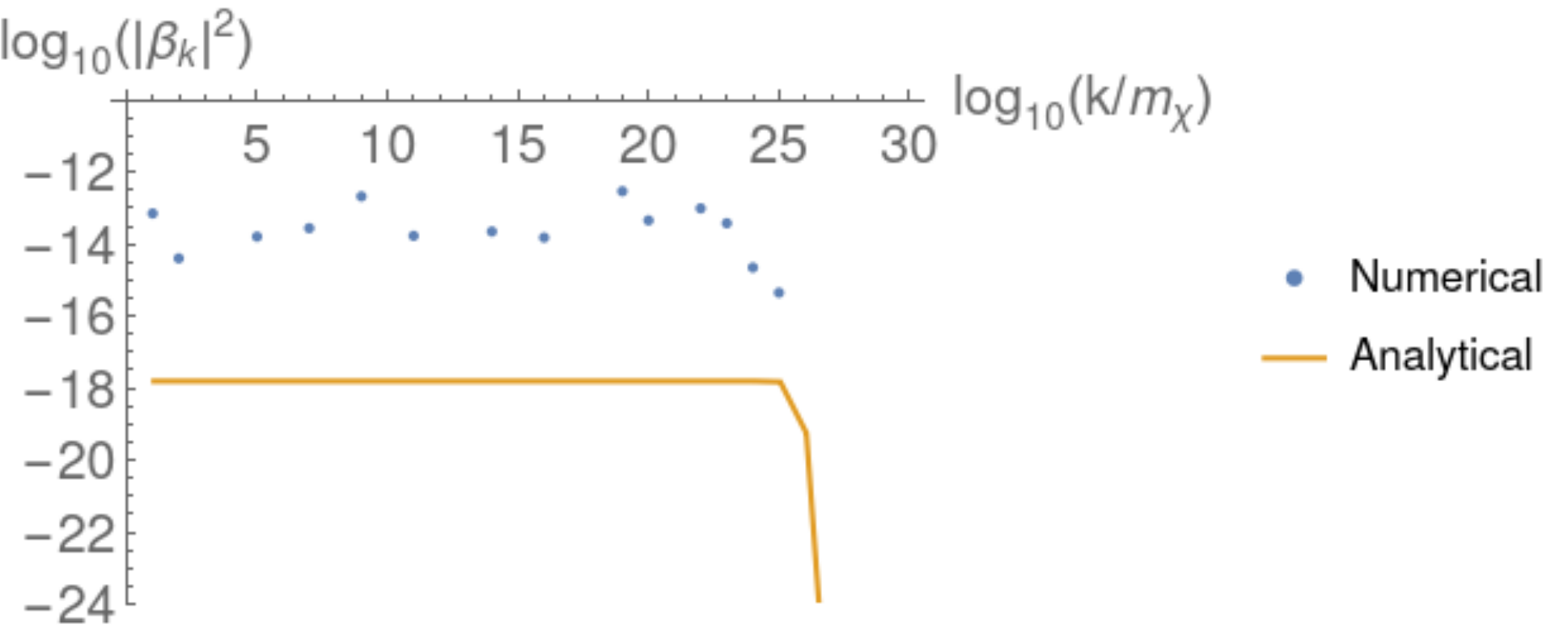}
\caption{Plot of the analytical and numerical values of $|\beta_k|^2$, corresponding to the model (\ref{quadratic}), for different values of $k$ obtained by solving equation \eqref{X} for $m_{\chi}=10^{15}$ GeV. We can see that there is no relation between both ways to calculate the $\beta$-Bogoliubov coefficient.}
\label{fig:beta_complete}
\end{figure}

\end{enumerate}

\section{Instant Preheating}

When  the reheating mechanism is due to the production of particles via Instant Preheating, as we have already seen,
the frequency is given by $
\omega_k(\tau)=
\sqrt{
k^2+a^2(\tau)[m_{X}^2+ g^2(\varphi-\varphi_{kin})^2]}$, and the 
standard way to obtain analytically the value of the $\beta$-Bogoliubov coefficient goes as follows:
First of all, we make the approximation 
$\varphi-\varphi_{kin,max}\cong \varphi_{kin}'\tau$, where we have chosen $\tau_{kin}=0$, and we assume a static universe with $a(\tau)=a_{kin}$. Then, the frequency becomes 
\begin{eqnarray}\omega_k^2(\tau)=k^2+a^2_{kin}[m_{X}^2+ g^2(\varphi'_{kin})^2\tau^2].\end{eqnarray}
Now we make the change of variable $\tau=\frac{y}{\sqrt{a_{kin}g\varphi'_{kin}}}$ to obtain the differential equation
\begin{eqnarray}
\frac{d^2\varphi_k}{dy^2}+(\kappa^2+y^2)\varphi_k=0,
\end{eqnarray}
where we have introduced the notation $\kappa^2=\frac{k^2+a_{kin}^2m_{X}^2}{a_{kin}g\varphi'_{kin}}$.

\

The positive frequency modes in the WKB approximation are
\begin{eqnarray}
\phi_{k,+}(y)=\frac{1}{(\kappa^2+y^2)^{1/4}}e^{-i\int \sqrt{\kappa^2+y^2}dy},
\end{eqnarray}
and for large values of $y$ one can make the approximation 
$(\kappa^2+y^2)^{1/4}\cong |y|^{1/2}$ and $\sqrt{\kappa^2+y^2}\cong |y|\left(1+\frac{\kappa^2}{2y^2}  \right)$, obtaining
\begin{eqnarray}
\phi_{k,+}(y\ll -\kappa)\cong |\tau|^{-1/2}\tau^{i\kappa^2/2}e^{i\tau^2/2}, \qquad \phi_{k,+}(y\gg \kappa)\cong |\tau|^{-1/2}\tau^{-i\kappa^2/2}e^{-i\tau^2/2},
\end{eqnarray}
and for the negative frequency modes one has 
\begin{eqnarray}
\phi_{k,-}(y\gg \kappa)=\frac{1}{(\kappa^2+y^2)^{1/4}}e^{i\int \sqrt{\kappa^2+y^2}dy}\cong |\tau|^{-1/2}\tau^{i\kappa^2/2}e^{i\tau^2/2}.
\end{eqnarray}

So, the frequency positive modes evolve as 
\begin{eqnarray}\label{evolution}
\phi_{k,+}(y\ll -\kappa)\longrightarrow \alpha_k\phi_{k,+}(y\gg \kappa)+\beta_k\phi_{k,-}(y\gg \kappa).
\end{eqnarray}

To obtain the Bogoliubov coefficients one can use the WKB method in the complex plane integrating the frequency along the path $\gamma=\{ z=|y|e^{i\alpha}, \pi \leq \alpha\leq 0\}$, obtaining that for $y\gg \kappa$ the early time positive frequency modes evolve at late time as
\begin{eqnarray}
e^{-\frac{\kappa^2}{2}\pi}|\tau|^{-1/2}\tau^{i\kappa^2/2}e^{i\tau^2/2},
\end{eqnarray}
and comparing with (\ref{evolution}) one gets
\begin{eqnarray}
|\beta_k|^2\cong e^{-\kappa^2\pi} \qquad \mbox{and}\qquad |\alpha_k|^2=1+|\beta_k|^2=1+e^{-\kappa^2\pi}.
\end{eqnarray}

Therefore, the
 energy density of the produced particles evolve as
\begin{eqnarray}
\langle\rho(t)\rangle=\frac{1}{2\pi^2 a^4(t)}\int_0^{\infty}\omega_k(t)k^2|\beta_k|^2dk\cong\frac{m_{eff}(t)(g\dot{\varphi}_{kin})^{3/2}}{8\pi^3}e^{-\frac{\pi m_{X}^2}{g\dot{\varphi}_{kin}}}\left(\frac{a_{kin}}{a(t)} \right)^3,
\end{eqnarray}
where $m_{eff}(t)=\sqrt{m_{X}^2+g^2(\varphi(t)-\varphi_{kin})^2}$ is the effective mass of the $\chi$-particles.

\

It is clear that this result was obtained by making the approximations we have depicted above, so it would be interesting to check if this result is also the same without performing these approximations and solving directly the system \eqref{Bogoliubovequation} using the methods described in the previous sections.

\section{The importance of $\beta_k$ in the calculation of the reheating temperature}
In this section we consider the gravitational particle production of superheavy particles (in that case one can safely ensure that the vacuum  polarization effects do not affect the inflationary period) and also Instant Preheating. In the second case, the bare mass of the produced particles $m_X$ is assumed to be negligible, but soon after its creation they acquire a very heavy effective mass which, as we have already seen in last section, is given by  
$m_{eff}(t)=\sqrt{m_{X}^2+g^2(\varphi(t)-\varphi_{kin})^2}$. So, in practice, in both mechanisms of reheating the masses are very heavy, and thus, in order to have a plasma of relativistic particles, they have to decay into lighter ones.
Then, two different situations will arise, namely the decay of these superheavy particles could be before or after the end of the kination period, and the reheating temperature differs in both cases.
\begin{enumerate}
    \item Decay before the end of kination: In the case of Instant Preheating, in order to evade an undesirable second inflationary period, it is mandatory that the produced particles decay into lighter ones before the end of kination \cite{fkl}.

    \
    
    First of all, it is important to realise that the energy density of the heavy particles scales  as $a^{-3}(t)$, and during kination the one of the inflaton field scales as $a^{-6}(t)$. Since during kination the energy density of the inflaton dominates, from the Friedmann equation one deduces that the Hubble rate scales as $a^{-3}(t)$ during kination. We recall as well that the decay finishes when the decay rate is of the same order as the Hubble rate, that is when 
    ${\Gamma}\sim H_{dec}=H_{kin}\left(\frac{{a}_{kin}}{a_{dec}} \right)^3$,
 where $\Gamma$ denotes the decay rate and the sub-index $dec$ means that the quantities are evaluated at the decay end.
 
   Then, the corresponding energy densities at the decay end will be
\begin{eqnarray}\label{LQIrho}
\rho_{\varphi, dec}=3{\Gamma}^2M_{pl}^2\qquad 
 \mbox{and}  \qquad
 \langle\rho_{dec}\rangle=
\langle\rho_{kin}\rangle\left(\frac{{a}_{kin}}{a_{dec}} \right)^3= \langle\rho_{kin}\rangle
\frac{\Gamma}{H_{kin}},\end{eqnarray}
where $\langle\rho_{kin}\rangle$ was already defined as the value of the energy density (\ref{vacuum-energy1}) at the beginning of kination, and it is the quantity that depends on the $\beta$-Bogoliubov coefficient.

Imposing that the end of the decay precedes the end of kination, which means $ \langle\rho_{dec}\rangle\leq \rho_{\varphi, dec}$, 
and taking into account that it is after the beginning of the kination,  i.e.,  $\Gamma\leq H_{kin}$,
one gets
\begin{eqnarray}\label{bound}
\frac{\langle\rho_{kin}\rangle}{3H_{kin}M_{pl}^2}\leq \Gamma\leq H_{kin}.
\end{eqnarray}

Finally,  the reheating temperature, i.e., the temperature of the universe when the relativistic plasma in thermal equilibrium starts to dominate, 
which happens when $\rho_{\varphi, reh}\sim \langle\rho_{reh}\rangle$, can be calculated as follows:
Since after the decay the evolution of the respective energy densities is given by
\begin{eqnarray}
\rho_{\varphi, reh}=\rho_{\varphi, dec}\left(\frac{a_{dec}}{a_{reh}}\right)^6 \qquad \mbox{and} \qquad
 \langle\rho_{reh}\rangle= \langle\rho_{dec}\rangle \left(\frac{a_{dec}}{a_{reh}}\right)^4, \end{eqnarray}
we will have
$\frac{ \langle\rho_{dec}\rangle}{\rho_{\varphi,dec}}=\left(\frac{a_{dec}}{a_{reh}}\right)^2,$
and thus,
from the Stefan-Boltzmann law $\rho_{reh}=\frac{\pi^2 }{30}g_{reh}T_{reh}^4$, where $g_{reh}=106.75$ is the effective number of degrees of freedom for the Standard Model, the reheating temperature will be
\begin{eqnarray}\label{reheating1}
 T_{reh}=  \left(\frac{30}{\pi^2g_{reh}} \right)^{1/4}
 \langle\rho_{reh}\rangle^{\frac{1}{4}} = 
 \left(\frac{30}{\pi^2g_{reh}} \right)^{1/4}
 \rho_{\chi,dec}^{\frac{1}{4}}
 \sqrt{\frac{\langle\rho_{dec}\rangle}{\rho_{\varphi,dec}}} \nonumber\\
\cong \left(\frac{10}{3\pi^2g_{reh}} \right)^{1/4} 
\frac{\langle\rho_{kin}\rangle^{3/4} }{M_{pl}^2H_{kin}^{3/4}\Gamma^{1/4}}
M_{pl},
\end{eqnarray}
which clearly depends on $\langle \rho_{kin}\rangle$, and thus on the $\beta$-Bogoliubov coefficient.

    \item Decay after the end of kination.

    When the decay of the $\chi$-field is after the end of kination (recall that kination ends when $\rho_{\varphi}\sim \langle\rho\rangle$),
one  has to impose ${\Gamma}\leq H(\tau_{end})\equiv H_{end}$, where we have denoted by $\tau_{end}$ the time at which kination ends. Taking this into account, one has 
\begin{eqnarray}\label{31}
H^2_{end}=\frac{2\rho_{\varphi, end}}{3M_{pl}^2}
\end{eqnarray}
and \begin{eqnarray}  \rho_{\varphi, end}={\rho}_{\varphi,kin}\left( \frac{{a}_{kin}}{a_{end}} \right)^6=
\frac{ \langle{\rho}_{kin}\rangle^2}{{\rho}_{\varphi,kin}},
\end{eqnarray}
where we have used that the kination ends when 
${ \langle{\rho}_{end}\rangle}={{\rho}_{\varphi, end}}$, meaning that
$\left({a}_{kin}/a_{end} \right)^3=
\frac{\langle{\rho}_{kin}\rangle}{{\rho}_{\varphi,kin}}$. So, the condition ${\Gamma}\leq H_{end}$ leads to the bound 
\begin{eqnarray}\label{bound1}
\Gamma\leq \sqrt{\frac{2}{3}}\frac{ \langle\rho_{kin}\rangle}{M_{pl}\sqrt{\rho_{\varphi,kin}}}.
\end{eqnarray}

\

 On the other hand, assuming once again instantaneous thermalization, the reheating temperature (i.e., the temperature of the universe when the thermalized plasma starts to dominate) will be obtained when all the superheavy particles decay, i.e. when $H\sim \Gamma$, obtaining
\begin{eqnarray}
T_{reh}=\left( \frac{30}{\pi^2 g_{reh}} \right)^{1/4}\langle\rho_{dec}\rangle^{1/4}= \left( \frac{90}{\pi^2 g_{reh}} \right)^{1/4}\sqrt{{\Gamma}M_{pl}},
\end{eqnarray}
where we have used that after  the end of the kination regime the energy density of the produced particles dominates the one  of the inflaton field. 

\

In that case the reheating temperature is only related with the decay rate $\Gamma$, which has to satisfy the bound (\ref{bound1}) which depends on $\langle\rho_{kin}\rangle$, and thus, on the $\beta$-Bogoliubov coefficient.

\end{enumerate}
\section{Conclusions}
The main goal of the present work is to point out the problems we have encountered  calculating numerically the $\beta$-Bogoliubov coefficient for smooth models. In fact,
we have presented several ways to calculate it numerically, but none of them seems to work correctly in the sense that we cannot ensure, for many reasons,  that the obtained results are the correct ones. For example, working with toy models where one can obtain an analytical solution, the numerical methods do not lead to the same results and, even worse, different numerical methods lead to different results.

\

Then, from our viewpoint, this is a problem that deserves complicated future studies at numerical level, because, as we have already shown,  the calculation of the $\beta$-Bogoliubov coefficient in non-oscillatory models is essential to obtain the reheating temperature of the universe. Maybe the problem is very technical and deep, and so,  numerical mathematics are needed, but our believe is  that it could be possible to get the right value of these coefficients with great accuracy.  

\section*{Acknowledgments} We would like to thanks Prof. Jaume Amor\'os for his valuable comments.
JdH is supported by grant MTM2017-84214-C2-1-P funded by MCIN/AEI/10.13039/501100011033 and by ``ERDF A way of making Europe'',
and  also in part by the Catalan Government 2017-SGR-247. L.A.S thanks the School of Mathematical Sciences (Queen Mary University of London) for the support provided.

\appendix

\section{Numerical methods}\label{app}

In this appendix we will discuss the way in which we have performed the numerical computations in order to calculate the Bogoliubov coefficient.

The procedure yields as follows:
\begin{enumerate}
    \item Solve numerically the continuity equation $\ddot{\varphi}+3H\dot{\varphi}+V_{\varphi}=0$ in cosmic time with initial conditions at the Hubble crossing.
    
    This arises no problems, since it leads to the expected results given in Fig $1$ in \cite{attractors}.
    \item Compute the scale parameter $a(t)=e^{\int_{t_0}^t H(s)}$. 
    
    Given that the value of the Hubble constant has been calculated numerically and $a(t)$ spans through more than $25$ orders of magnitude, this is something that needs to be done more accurately. So far, we have tried to obtain $a(t)$ both by computing the above integral or as the solution of the simple ODE $a'(t)=H(t)a(t)$.
    
    \item Once having $a(t)$, solve the ODE \eqref{B} or \eqref{X} for the different range of values of $k$ in order to obtain the value of the Bogoliubov coefficient. 
    
    We have used the ordinary NDSolve function in Mathematica trying to tun the needed precision and accuracy { as well as the Filon-type methods in Matlab}, but all have happened to lead to different results for the two sets of ODEs. 
\end{enumerate}

Our guess is that the step $2$ has not been successfully achieved and we are using wrong values of $a(t)$ in step $3$.


\begin{thebibliography}{99}

\bibitem {guth}
A. Guth,
Phys. Rev. {\bf D 23}, 347 (1981).



\bibitem{riess}
  A. G. Riess et al,
  Astron. J. {\bf 116}, 1009 (1998) [arXiv:astro-ph/9805201].
  
  
  \bibitem{perlmutter}
    S.  Perlmutter  et  al,
    Astrophys.  J. {\bf 517},  565(1999) [arXiv:astro-ph/9812133].

















\bibitem{pv}
P. J. E. Peebles and  A. Vilenkin,
Phys. Rev. {\bf D 59}, 063505 (1999)  [arXiv:astro-ph/9810509].





\bibitem{parker}
L. Parker, 
Phys. Rev. Lett. {\bf 21}, 562 (1968).

\bibitem{ford}
L. H. Ford, 
Phys. Rev. {\bf D 35}, 2955 (1987).


\bibitem{fkl0}
G. Felder, L. Kofman and  A. Linde, 
 	Phys. Rev. {\bf D 59}, 123523 (1999)  	[arXiv:hep-ph/9812289]




\bibitem{fkl}
G. Felder, L. Kofman and  A. Linde,
Phys. Rev. {\bf D 60}, 103505 (1999) [arXiv:hep-ph/9903350].



\bibitem{damour} 
T. Damour and A. Vilenkin,
Phys. Rev. {\bf D 53},  2981 (1996) [arXiv:hep-th/9503149].

\bibitem{giovannini} M. Giovannini,
 Phys. Rev. {\bf D 58}, 083504 (1998)
[arXiv:hep-ph/9806329].







\bibitem{Bunch}
 T. S. Bunch, 
J. Phys.  {\bf A 13}, 1297 (1980).




{ \bibitem{Zeldovich}
Ya. B. Zeldovich and A. A.
Starobinsky, JETP {\bf 34}, 1159 (1972).}

\bibitem{gmmbook}
A. A. Grib, S.G. Mamayev and V. M. Mostepanenko,
Friedmann Laboratory Publishing for Theoretical Physics,
St. Petersburg (1994).


\bibitem{ema}
Y. Ema, K. Nakayama and  Y. Tang,
 JHEP {\bf 09}, 135 (2018) 	[arXiv:1804.07471 [hep-ph]]. 
 

\bibitem{hashiba}
S. Hashiba and J. Yokoyama, JCAP {\bf 01}, 028 (2019) 
	[arXiv:1809.05410 [gr-qc]].


\bibitem{haro}
L. Arest\'e Sal\'o, D. Benisty, E. I. Guendelman and J. de Haro,
Phys. Rev. {\bf D 103}, 123535 (2021)
	[arXiv:2103.07892 [astro-ph.CO]].





\bibitem{haro1}
L. Arest\'e Sal\'o and J. de Haro,
Phys. Rev. {\bf D 104}, 083544  (2021)	[arXiv:2108.10795 [gr-qc]].

\bibitem{haro2}
L. Arest\'e Sal\'o and J. de Haro,
	[arXiv:2112.12992 [gr-qc]].
	
	\bibitem{attractors}
	L. Arest\'e Sal\'o, D. Benisty, E.I. Guendelman and J. de Haro,
	Phys. Rev. {\bf D 103}, 123535 (2021) [arXiv:2103.07892 [astro-ph.CO]]

\end{thebibliography}
\end{document}